\documentclass[12pt, a4paper,superscriptaddress,groupedaddress]{revtex4}   
\usepackage{amsmath}
\usepackage{bbm}
\usepackage{graphicx}  
\usepackage{dcolumn}  
\usepackage{bm}        
\usepackage{amssymb}   
\usepackage[section]{placeins}
\usepackage{bbm}
\usepackage{float}

\newcommand{\q}[1]{``#1''}

\begin{document}

\title{Hierarchical searching in episodic memory} 

\author{Francesco Fumarola}
\noaffiliation

\date{\today}
 
\begin{abstract}   
An analysis of free-recall datasets from two independent experiments allows to identify two anomalous instances of non-monotonicity in free recall: a maximum in the dependence of the inter-response intervals on the serial-position lags, and a minimum in the rate of contiguous recall near the beginning of the recall process. Both effects, it is argued, may stem from a hierarchical search protocol in the space of memories. An elementary random-walk model on binary strings is used to test this hypothesis. 
\end{abstract}     
           
\maketitle

\section{Introduction}
 
Free-recall experiments are a key tool for the controlled investigation of episodic memory. A typical free-recall experiment takes place in two stages: during the presentation stage, subjects are shown a list of words; during the memory test, they are requested to recall them in any order  (Binet and Henri, 1894; Murray, 1960).

It will be useful to introduce the following terminology: \q{serial position}, the numbered position of a word within the list; \q{inter-response interval} (IRI), the time interval that elapses between two consecutive recalls; \q{transition}, any sequence of two consecutively recalled word; \q{lag}, the difference between the serial positions of two words in a given transition. For example, if the $5$th word in the list is recalled right after the $8$th, the corresponding lag is $L= -3$. Transitions with lag $L=\pm 1$ will be called contiguous. 

Some of the best-known empirical properties of free recall are: 
\begin{itemize}
  \item 
 power-law scaling: the number of retrieved items scales like a power law of the number of items in the list  (Standing, 1973; Murray et al., 1976); 
\item primacy and recency effects: the first and last words in the list are recalled better than the rest  (Murdock, 1962);
   \item
   the lag-recency effect: the probability of transitions from one recall to the next is a decreasing function of the lag (Kahana, 1996). 
\item 
   forward asymmetry: the tendency to recall items in forward order (Ebbinghaus, 1913).
\end{itemize}
 
Many other effects have been reported, especially in connection with the semantic and phonological properties of individual words (for a review, see Kahana, 2012).
 
In this paper, two additional effects are pointed out, concerning the dependence of the IRIs on the lag and the time-dependence of the contiguous recall rate.
 
I begin by considering, in the next section,  two datasets coming from independent experiments and by extracting the variables I have mentioned. The effects emerge clearly from both datasets, despite rather different experimental conditions. I then sketch qualitatively a possible explanation -- the presence of a hierarchical component in the search mechanism. I design a simple model that may help test the hypothesis; the model is a variant of the random-walk approaches to free recall recently adopted by various authors.  Finally, I compare the findings with the result of simulations on the model and consider possible ramifications of the hypothesis. 
 
\section{Non-monotonicity in Free-Recall Observables}
  
\subsection{The Archival Data}
  
The PEERS study (Penn Electrophysiology of Encoding and
Retrieval Study) is a study recently conducted at the University of Pennsylania and devoted to assembling a large database on the electrophysiological
correlates of memory (Lohnas, 2013).
 
The sample I have considered corresponds to 
Experiment I of PEERS. It includes data from trials on $156$ college students (age range: 18$-$30) and on $38$ older adults (age range:
61$-$85 years). In each trial, $16$ words were presented one at a time on a computer screen. Each word was drawn from a pool of $1638$ words with heterogeneous semantic and lexical properties.

Each item was kept on the screen for $3000$ ms,
followed by an interstimulus interval of
800$-$1200 ms. After the last item in the list, there was a delay of 1200$-$
1400 ms, after which the participant was given $75$ s to attempt to recall aloud any of the just-presented items. Multiple trials were performed on each subject, summing up to $3744$ trials for the students' sample and to $912$ for older adults. For more details on the experimental procedure, see 
Healey and Kahana (2016).

In order to check my findings against multiple experiments, I have also made use of an older set of data, coming from the free-recall experiments described in Polyn et al., 2009.

These experiments involved 45 participants; the lists contained 24 words each, again presented on a computer screen. Each word was shown for 3 s; in the retrieval stage, participants were given 90 s to recall the words. A total of 1394 trials were performed. 
 
Three differences may be noticed between the experimental conditions of PEERS and those employed by Polyn et al.: (1) the length of the lists, which is $16$ in PEERS  and $24$ in Polyn et al.; (2) the longer time allowed by Polyn et al. for both memorization and recall; (3) the different size of the word-pools, as PEERS experiments used a pool of 1638 words, while the pool of Polyn et al. contained only 1297.    
 
Intrusions (i.e. words wrongly recalled from outside the list) have been discarded from all these data; repetitions (less than $0.3\%$ of events) are counted with the lag $L=0$. 
 
\subsection{Distribution of the IRIs}
 
The free-recall literature has shown, as mentioned, that two events are more likely to be recalled at a short distance from each other if the serial-position interval between them is smaller (lag-recency effect). There is therefore a degree of chronological continuity in the way events are stored and recalled, and this fact has been successfully formalised in retrieved-context theories such as the Temporal Context Model of Howard and Kahana, 2002.
    
Accordingly, we expect the $\textit{time}$ $\textit{interval}$ between two $\textit{consecutive}$ recalls to be also an increasing function of the time interval between the events recalled. Otherwise said, we expect that if two events have occurred at a longer time distance from each other, it will take a longer time to recall them consecutively. In the specific case of free-recall experiments, the inter-response interval (IRI) between two consecutive recalls should then be an increasing function of the absolute value $|L|$ of the lag between the words recalled.

This simple expectation turns out to be contradicted by the data. To see this, let us begin the analysis with a rough approximation, i.e. by regarding all recorded transitions as statistically independent.  The resulting histograms of the IRIs for the experiments of Healey and Kahana (2016) are displayed in the upper-left panel of Fig. 1. A different histogram is obtained for each absolute value of the lag; to avoid cluttering, some intermediate lag values are not shown; since the lists in these experiments were $16$ words long, the maximum lag is $15$. 
 
\begin{figure}
\includegraphics[width= \textwidth]{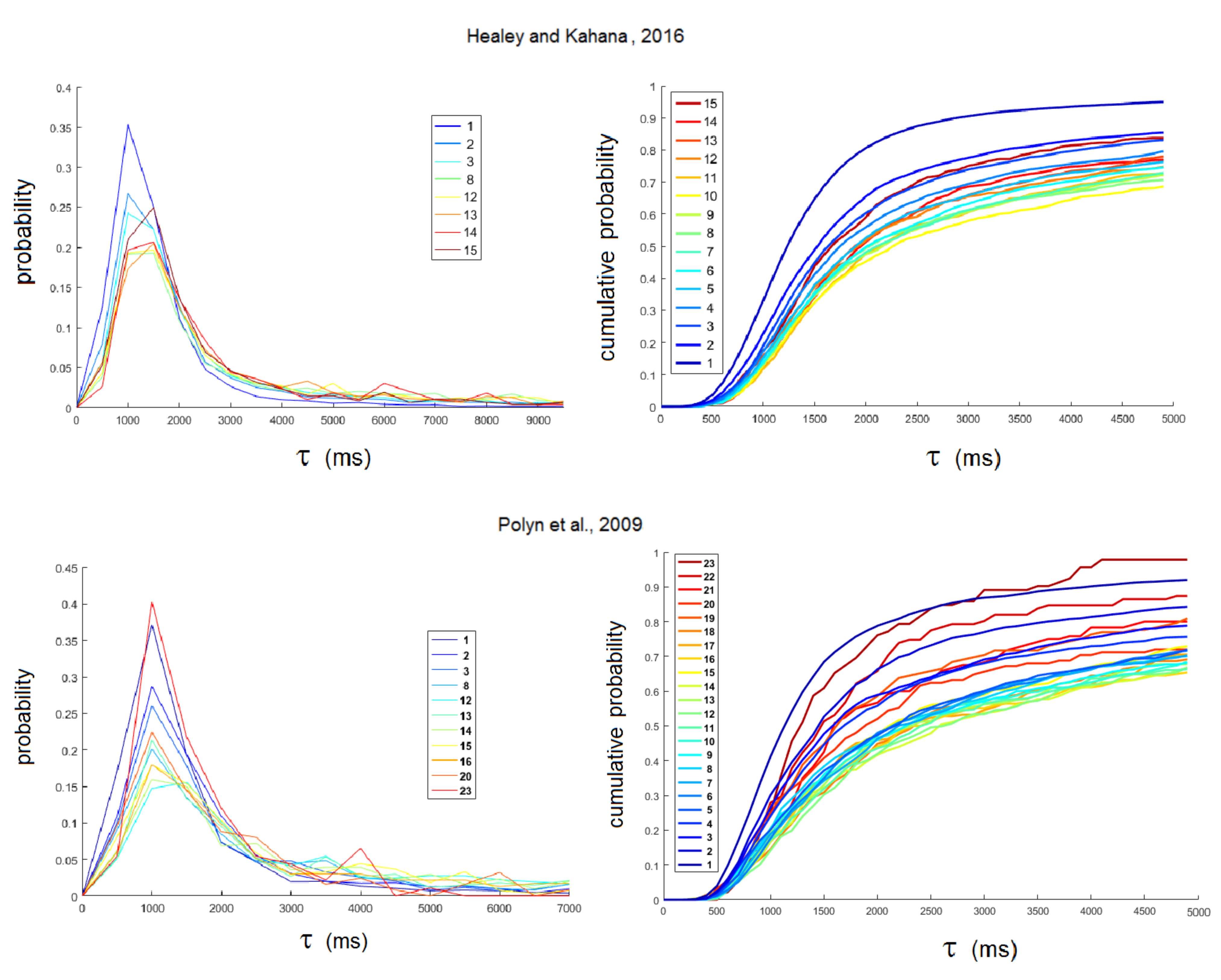}
\caption{Inter-response intervals in the experiments of Healey and Kahana (2016) and Polyn et al., 2009. Left-hand panels: distribution of IRIs for representative values of the lag. Right-hand panels: cumulative probability for all lags. Values of the lag are shown in the legend.}
\centering 
\end{figure}

The presence of a peak at small $\tau$ for transitions with  $|L|=+1$ is no surprise, and agrees with expectations from the lag-recency effect. For longer lags, the peak is suppressed, as seen in the curves corresponding  $|L|=2,3,8$. However, this monotonous suppression has ceased by $|L|=12$, and the peak has begun to grow again. With the maximal lag $|L|=15$, we have recovered a peak as tall as the peak for $|L|=3$. This means that the recall process, slow for transitions to intermediate lags, is faster again for transitions to very long lags. 
    
The upper-right panel shows the corresponding cumulative distribution, visualised for all values of the lag. The growth of the cumulative toward saturation becomes gradually slower as the lag increases from $|L|=+1$. It reaches its slowest point for $|L|^*=10$ (yellow curve); then it gradually grows faster again. The inversion point is the lag $|L|^* \sim 0.6 S$, where $S$ is the size of the lists. For the maximal lag $|L|=15$, the cumulative curve virtually overlaps with the curve for $|L|=3$. 

Data from the experiments of Polyn et al. (2009) are similarly displayed in the two lower panels.  The lower-left panel shows the distribution of IRIs for a handful of lag values. The behavior of the distribution confirms what we observed in the data of Healey and Kanaha: a suppression of the peak for intermediate lags values, and a new enhancement for near-maximal values. In fact, we find here that the peak for the maximal lag (that is, in this case, $|L|=23$) is taller than the peak for contiguous transitions ($|L|=1$).   
    
These findings are better gauged by plotting again the cumulative distribution, shown in the lower-right panel of Fig. 1 for all lags. It can be seen that the highest suppression of the peak occurs for $|L|^*=14$, so we have once again $|L|^* \sim 0.6 S$, where $S$ is the size of the lists. For the maximal lag $|L|=23$, the cumulative curve reaches saturation faster than the curve for contiguous transitions. 
     
While this is an intriguing result, it relies on the assumption that all transition events could be treated independently. On the other hand, transition events within the same trial are statistically correlated, and the same may be true for transition events within different trials performed on the same subject.     
         
To better compare these sets of data, I have averaged the IRIs corresponding to all transitions performed both on the same subject and with the same value of the lag. In Fig. 2, results are plotted as functions of the ratio between the lag and the size of the lists. Blue dots correspond to individual subjects, black curves are the corresponding histograms, while the red and green curves display the mean and median over the histograms. 
   
Participants in the experiment of Polyn et al. were accorded a longer time for the memory test. Perhaps, this is why the curve of the mean IRI for Polyn et al. lies slightly above the curve for the data of Healey and Kahana. In spite of this, the two curves follow the same overall pattern.

\begin{figure}
\includegraphics[width= \textwidth]{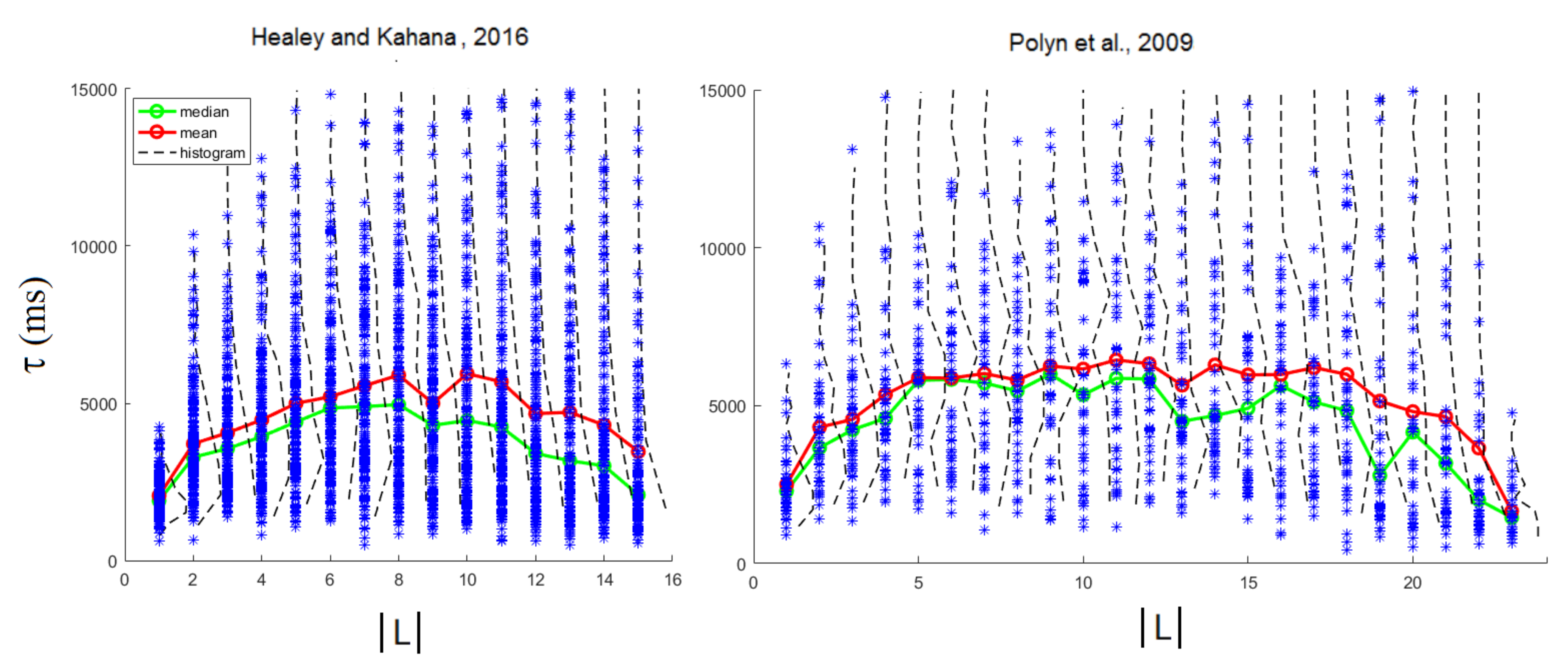}
\caption{Mean inter-response interval for transitions with a given lag, computed from the data in Healey and Kahana, 2016 (left-hand panel) and in Polyn et al., 2009 (right-hand panel). Each blue dot corresponds to a different subject, the black curves show histograms for each lag, and green/red curves are medians/means over all subjects.}
\centering 
\end{figure}

For short lags, a growth in the mean IRI is observed. This requires no explanation, as we expect the thought process to move with greater speed between memories created within a shorter time from one another. As the lags grow, so does the time it takes to go from one memory to the next. This increasing trend, however, slows down in mid-range, and ceases altogether at values of $|L|/S$ in the range between $0.5$ and $0.7$. Then the average IRI begins to decrease as the lag grows.  

I have computed the correlation coefficients for the ascending and descending arcs of the two histograms, by regarding the maximum as their divide point. The ascent is characterised by $r =0.31$ in Healey and Kahana, by $r= 0.32$ in Polyn et al.; the descent, by $r= - 0.18$ in Healey and Kahana and $r = - 0.23$ in Polyn et al. All these four coefficients correspond to a p-value $p<10^{-5}$.

\begin{figure}
\includegraphics[width= \textwidth]{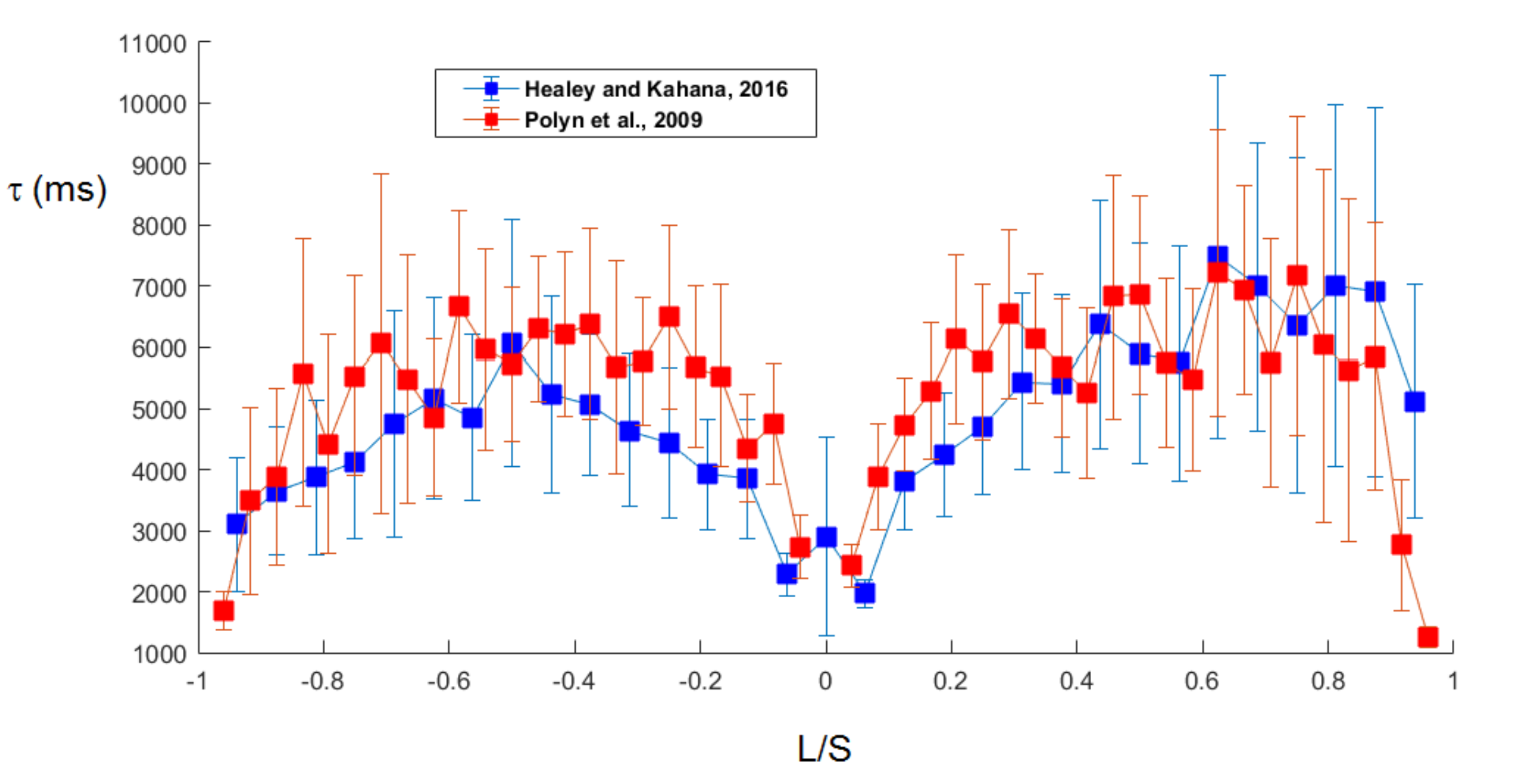}
\caption{Mean IRIs as a function of the ratio between lag and list size, counting the lag signs. Error bars are computed over the subject-by-subject distributions. The non-monotonous behaviour survives for both positive and negative lags. 
}
\centering 
\end{figure}
The decreasing component of these curves is a highly counterintuitive feature. It suggests that, if two events take place within a shorter time interval, they will be recalled further apart from each other (at least for a certain range of time intervals). The Polyn et al. data are particularly surprising, because the curve reaches further down for maximal lags than for minimal lags. A memory situated $23$ memories away from the most recently visited memory is recalled faster, on average, than a memory contiguous to the last one we recalled.

It is well known that transitions with a positive lag have a higher probability of occurring than transitions with a negative lag. Since the free-recall statistics is not symmetric as a function of the lag, one may wonder if the phenomenon we have just discovered concerns only negative transitions, or only positive ones. 

In Figure 3, the answer is sought by plotting the mean IRIs for both sets of experiments. The means were computed by averaging values concerning individual subjects, and the error bars are their standard deviations. The two curves are again nearly overlapping, in spite of differences in the two experimental settings. 

More importantly, the non-monotonous pattern is identically displayed by both curves on both sides of the origin, with the increasing trend followed by a decreasing section. The asymmetry can be  measured by computing the index $\eta= \frac{1}{S -1} \sum_{n=1}^{S -1 } \frac{\left|\bar{\tau}(n) - \bar{\tau}(-n)\right|}{\tau(n) + \tau(-n)}$ for the two sets of data, which gives $\eta_{2009}=   0.08 $
and $\eta_{2016} =   0.12$. The psychological mechanism at the root of this behavior, we may conclude, is likely to be independent of forward asymmetry.
   
\subsection{Contiguous Recall Rate}
 
Let $t$ be the descrete time variable labeling steps in the recall process. The contiguous recall rate $p_{cont}(t)$ at the $t$-th step is the probability that the $t$-th recall will be effected contiguously. Otherwise said, $p_{cont}(t)$  is the probability that the serial position of $t$-th word recalled will be contiguous to the serial position of the $t$-th word recalled. 

How do we expect $p_{cont}(t)$ to evolve as a function of time? We will first seek a tentative answer under some simple assumptions, then look at the actual answers. 
 
In Fig. 4, I have plotted the distribution of the initial recall probability, as a function of serial position, for the two datasets under consideration. This is the probability distribution $p_1$ of the serial position of the first word to be recalled, and we will refer to this distribution as the \q{initial conditions} of the process. It has been studied extensively 
(Murdock, 1960; Murdock, 1962; Bjork and Whitten, 1974) and, as per the primacy and recency effects, it is concentrated near the beginning and the end of the list.  

\begin{figure}
\includegraphics[width= \textwidth]{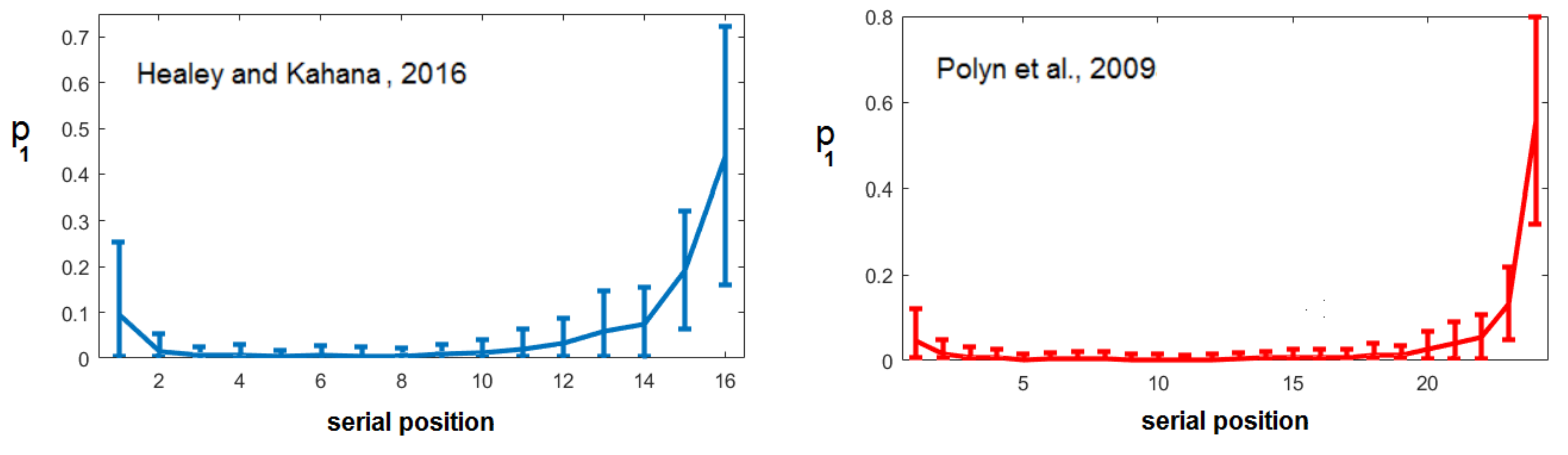}
\caption{Initial-recall probability as a function of serial position, for the two experiments. Error bars refer to the statistics over different subjects.}
\centering 
\end{figure}

The time-averaged probability $P^0_{j i}$ of transitions from word $i$ to word $j$ is shown in figure 5A, computed for the case of the Polyn et al. data. These probabilities have been obtained by considering transitions recorded at any point during the recall process and performing an unweighed average over events. Therefore, they do not contain $\textit{all}$ the information about the recall process unless the latter is a homogenous Markov chain. 
 
Immediate repetitions, nearly absent, have not been considered, hence the diagonal is zero. The super- and sub- diagonal elements are the most prominent terms. The transition probability decays quickly as we move toward the off-diagonal corners. Notice that there is always a finite probability for recall termination, shown as an additional column in the plot.  This is equivalent to the presence of a sink state in the graph associated to the process. 
 
We can now apply the matrix $\hat{P}^{0}$ to the distribution $p_1$ in order to obtain stochastic realizations of the supposed Markov process, over which we may compute the statistics of arbitrary variables. In particular, we are interested in the variable $P_{cont}(t)$, the contiguous recall probability, defined above. 

\begin{figure}
\includegraphics[width= \textwidth]{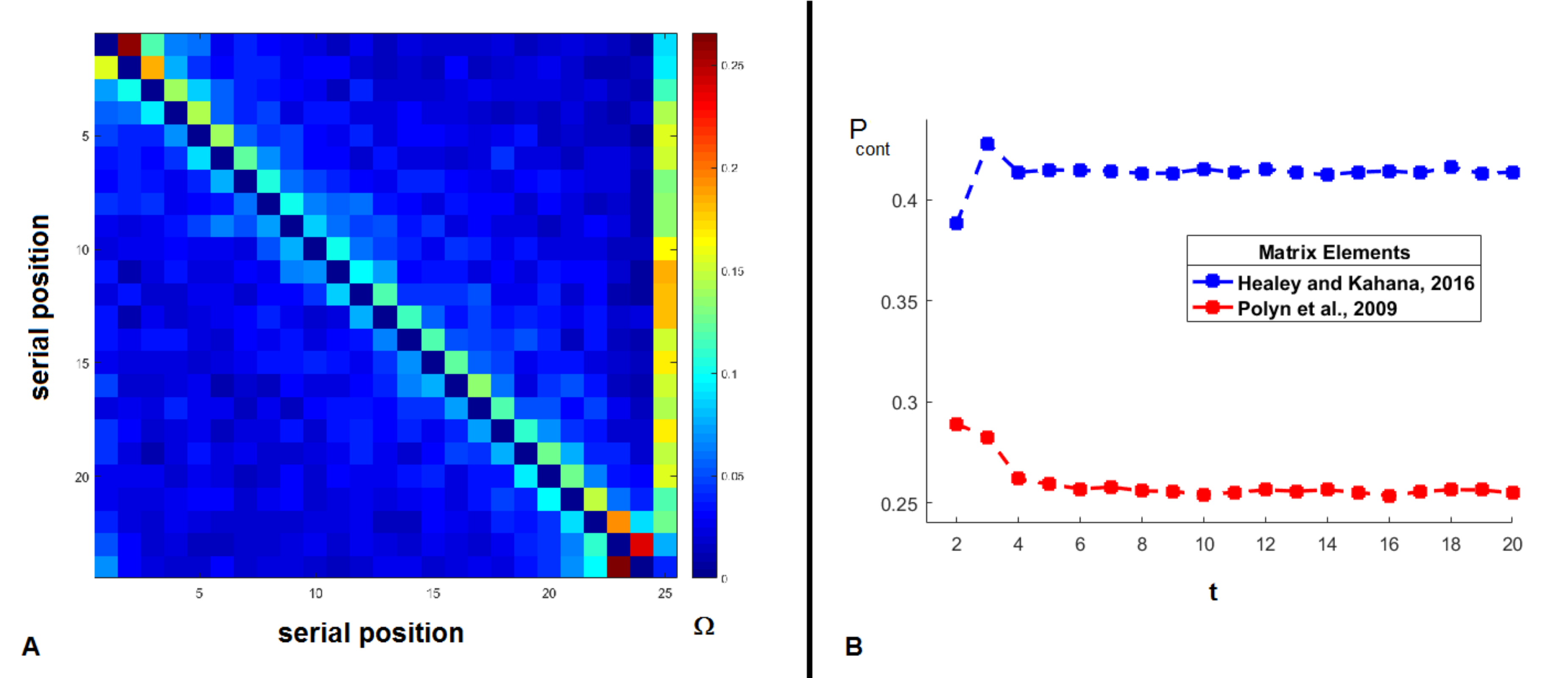}
\caption{Panel A: Matrix of time-averaged transition probabilities between serial  positions for the data of Polyn et al., 2009; color-coding is illustrated by the sidebar; the sink state $\Omega$ indicates recall termination. Panel B: probability of contiguous recall as a function of the order of recall, computed markovianly from the Polyn et al. initial conditions with the transition matrix shown in panel A, and from the Healey and Kahana initial conditions with the equivalent transition matrix .}
\centering 
\end{figure}
  
Results are shown in panel B of figure 5. A large variability in the contiguous-recall probability emerges at the very beginning of the recall process. This is due, of course, to the fact that the initial distribution is distant from the equilibrium state. The two curves approach stationary values rather quickly. The stationary value of the contiguous recall probability has no dependence on the distribution $p_1$; it is higher for the Healey and Kahana data because, in those experiments, shorter lists have been used, hence contiguous transitions are statistically favored. 
  
The homogenous Markovian hypothesis tells us that a quick relaxation to a steady value is the main feature of the contiguous recall rate $p_{cont}(t)$. A direct analysis of data, however, yields a completely different picture, in which the most conspicuous feature is a \q{minimum} in contiguous recall near the beginning of the recall process. 
    
This is shown in Fig. 6, where the contiguous recall rate has been extracted by analyzing transition events in the two sets of experimental data. The blue dots are probability values obtained by averaging over all trials corresponding to the same subject. Black curves are histograms corresponding to each given lag, while the red curve is the average of the probability values corresponding to each subject. 

\begin{figure}
\includegraphics[width= \textwidth]{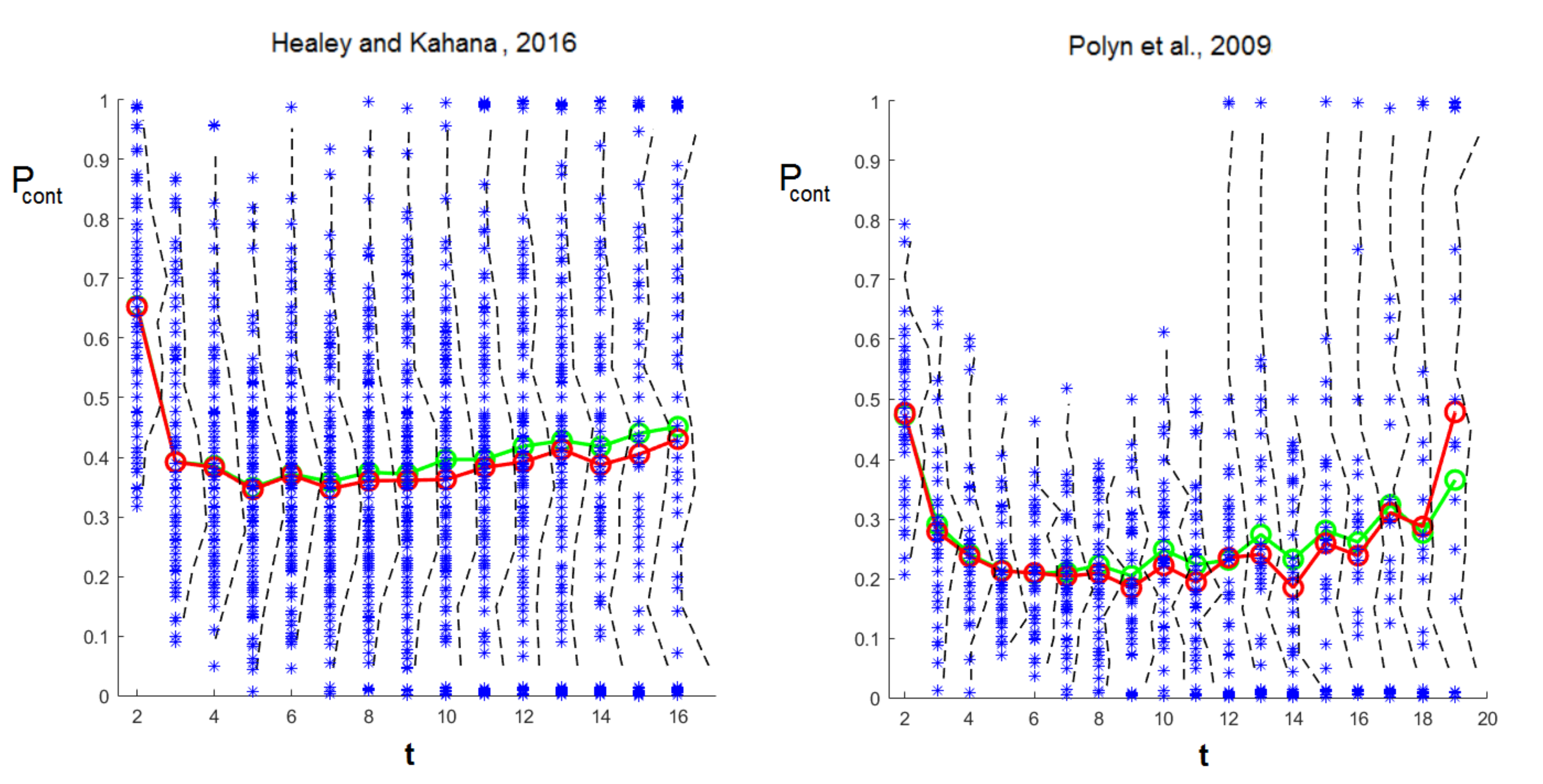}
\caption{Rate of contiguous recall as a function of the order of recall, computed over the two datasets.  Blue dots: probability value corresponding to individual subjects; black curves: histograms corresponding to each given lag; red curve: mean of the probability for ecah subject, averaged over subjects; green curve: probability computed by regarding all trials as independent.}
\centering 
\end{figure}

As can be seen, both simulations and experiments show a large variability in the contiguous-recall probability at the very beginning of the recall process. This is due, at least in the simulations, to the fact that the initial distribution is distant from the equilibrium state. However, unlike the simulated curve, the experimental curve experiences a minimum at a recall position $t^*$, such that $t^* \sim 5$ for 16-word lists, and $t^* \sim 9$ for 24-word lists. 

The contiguous recall rate then enters an increases phase, while the simulated curve stays close to a stationary value. For very large times, the experimental curves are no longer reliable as too few experimental points are available. 
  
For comparison, the average probabilities obtained by regarding all trials as independent have been included as the green curve in Fig. 6. The behavior of the trial-averaged and subject-averaged curves is analogous, even if a different number of transitions is recorded for each subject. Just as we found for the IRI distribution, the subject-by-subject and trial-by-trial statistics are in qualitative agreement.  

I have finally computed correlation coefficients for the descending and ascending branches of the contiguous recall curve. Between the beginning of the recall process and the minimum of the curve,  the correlation coefficient is found to be $r=- 0.42$ for the Healey and Kahana data and $r=-0.47$ for the Polyn et al. data. Both these correlation coefficients correspond to $p<10^{-5}$. From the minimum to the last experimental point in the plot, correlation coefficients are found to be $r=0.10$ for Healey and Kahana, $r=0.22$ for Polyn et al., with $p < 10^{-4}$ in both cases. 
  
Once again, we were expecting a monotonous trend and discovered a non-monotonous one. While in the IRI distribution we found the presence of a maximum, in the contiguous recall rate we detected a minimum. These two counterintuitive phenomena may, of course, result from independent mechanisms. In the second half of this paper, I will develop a particular hypothesis that may be a viable candidate to explaining both effects. 

\section{Hierarchic Search Hypothesis}

\subsection{Hypothesis and toy model} 
    
A natural approach to modeling free-recall relies on processes exploring stochastically a psychological space populated by available memories. This corresponds for instance to the model of memory retrieval used in Kenett (2014) or Abbott (2015),  and it could be argued that conventional retrieved-context models are based on a similar principle, because temporal contexts are encoded by a matrix representation that includes a stochastic element. 
  
Romani et al. (2013) represented the thought process as a random-walker moving in a particular geometry and were able to approximate, under appropriate conditions, more complex neural-network models of associative retrieval, predicting correctly the power-law scaling of free recall that we mentioned above.

In the random-walker scenario, however, it would be difficult to understand either of the two effects described in the previous section. In particular, the average IRI would have to be a monotonously increasing function of the lag. This follows from the fact that the distance between two memories is larger if the events memorised happened further apart in time. But the lag-recency effect entails that the random-walker moves continuously in space, hence a longer distance will be covered on average in a longer time. Therefore, models based on a random walker are not likely to explain why the average inter-response interval $\textit{decreases}$ as the lag grows beyond a certain point.

On the other hand, there has been a lot of work in recent times regarding the brain's ability to carry out concurrent thought processes simultaneously 
(Sigman and Dehaene, 2008;
Cowan, 2010); in computer science, studies have emerged on the advantages of carrying out certain tasks through a swarm-like organization of computational components (Bonabeau et al, 1999; 
Eberhart, 2001). This motivates the question of how the random-walker envisaged by Romani et al. (2013), 
Kenett (2014), Abbott (2015) would behave if replaced by several random-walkers simultaneously searching the psychological space. 
  
The possibility we will focus upon consists in a hierarchical search, where multiple random-walkers move simultaneously in search of memories and the right to effect retrieval resides with one random-walker at a time. The space in which the random-walkers move will be modeled, for the sake of simplicity, as the space of binary arrays where each point corresponds to a string of $0$'s and $1$'s; the length of the strings is the dimensionality $d$ of the space.

The presentation stage of free recall will be modeled as the process that, starting with a given binary array, creates new memories by progressively flipping a random digit. The constraint that no state should correspond to two memories will be enforced, although in sufficiently high dimensions it is mostly unnecessary.  
  
At the beginning of the retrieval stage, each random-walker is made to lie at the location of one of the memories. We pick these initial locations randomly and independently of each other, according to the distribution of initial words found in experimental data (Fig. 4). The retrieval process for each individual walker evolves by flipping random digits in the binary array that describes the walker's location. Whenever the location coincide with a memory, that memory is a candidate for retrieval. 
 
The right to effect retrieval reside on one random-walker at a time, and only that walker is allowed to retrieve memories. This scenario is hierarchical (fig. 7) because it combines two stochastic motions: the \q{microscopic} motion of each individual random-walker, representing parallel thought-tracks, and the \q{macroscopic} motion of the retrieval rights, switching stochastically among them. The latter will be assumed to be self-avoiding,  no random-walker being visited twice during the recall process.

The number $N$ of walkers deployed for recall may be expected to be proportional to the magnitude of the recall task, that is, to the length $S$ of the list.  Random-walkers that the macroscopic walk has not yet visited will be called \q{unexplored}. If $n_t$ is their number at a generic time step $t$, the system has a finite probability $q(n_t)$ of hopping onto any of the unexplored random-walkers, and a probability $1 - n_t q(n_t)$ of staying with the current one.  Once a random-walker has been abandoned, it is no longer be taken under consideration.

\begin{figure}
\includegraphics[width= \textwidth]{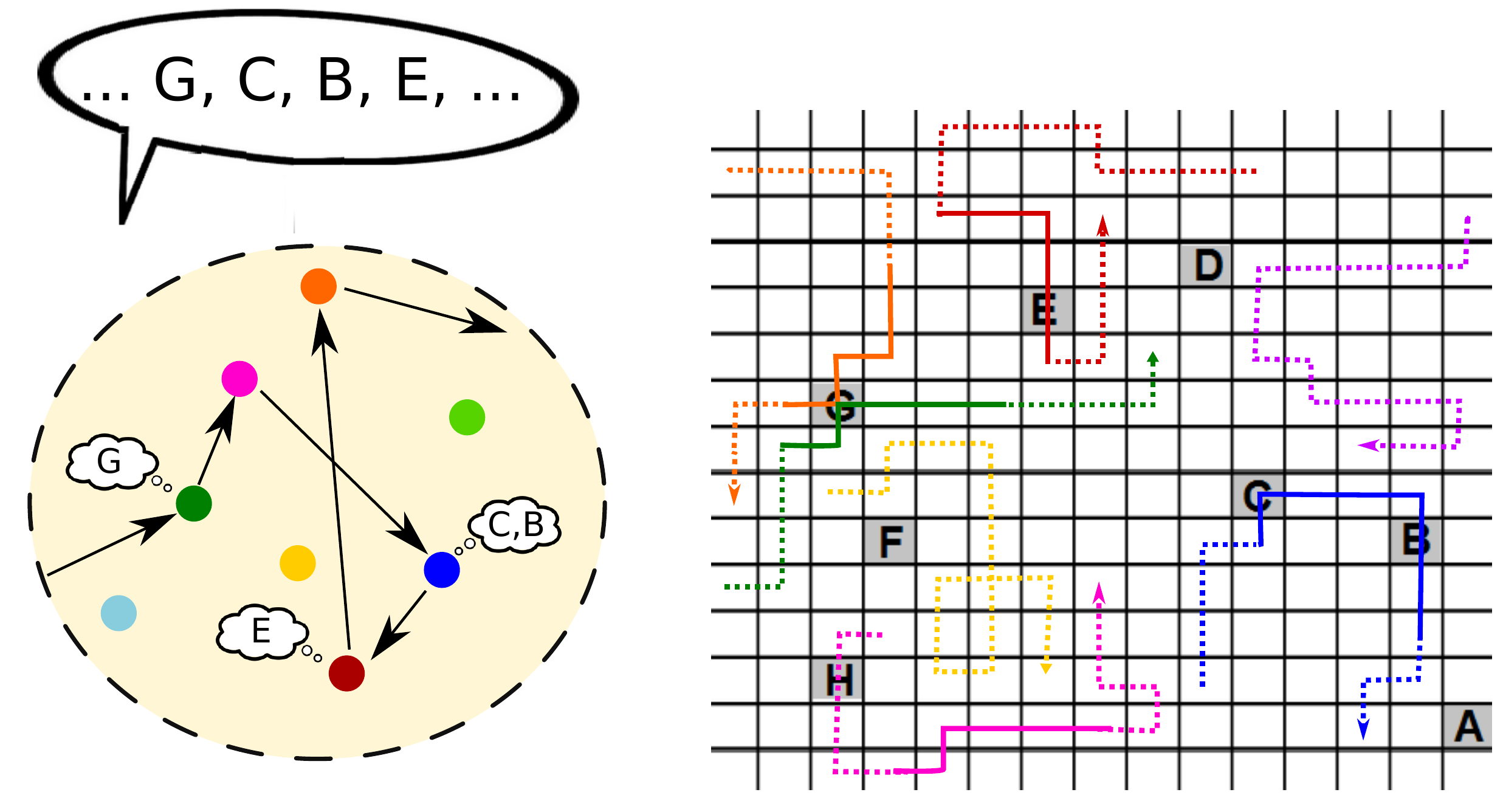}
\caption{Example of dynamics in the hierarchical model. Left-hand panel: random-walk on the graph of random-walkers, with corresponding recall events. In the example, only word B is recalled associatively. The rest are located by switching among random-walkers. Right-hand panel: cartoon depiction of psychological space. Word-memories are represented as gray cells; stretches of the trajectory during which a random-walker is endowed with retrieval privileges are shown as solid lines, the rest as dotted lines. }
\centering 
\end{figure}

\subsection{Numerical results} 
  
Results from simulations of this model are shown in Fig. 8, for the simple choice $ q(n)=\frac{1}{n +1}$. At the beginning of each simulation, each random-walker has been placed at the location of a word-memory, chosen in accordance with the empirical distributions in Fig. 4. The model has two parameters: the dimensionality $d$ of psychological space and the walker-per-word ratio $\kappa$, defined by $N=\kappa S$; the results in Fig. 8 were obtained in $7$ dimensions and for $\kappa=9$.
  
\begin{figure}
\includegraphics[width= \textwidth]{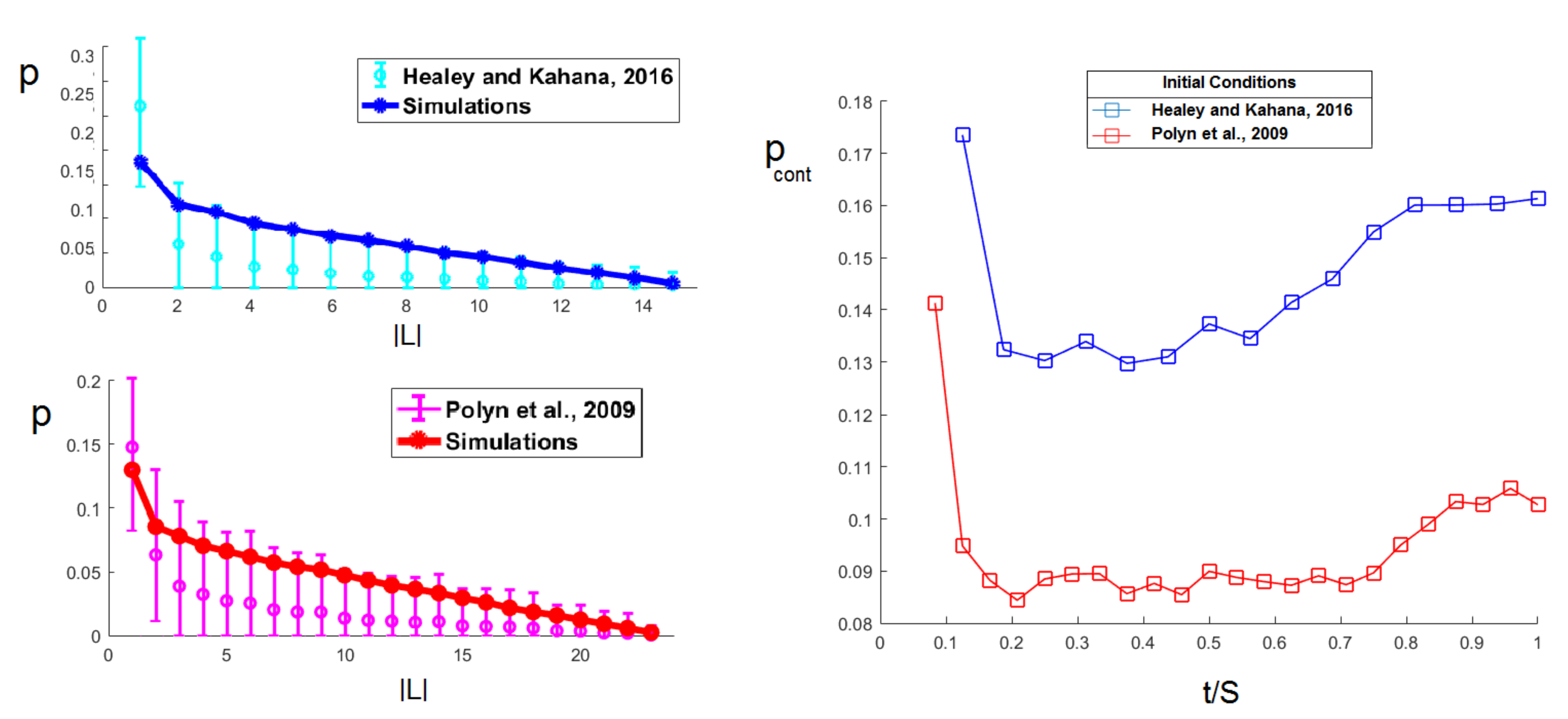}
\caption{ Simulation of the memory search. Left-hand panel: recall probability as a function of lag, computed over individual subjects and averaged for the two datasets, and compared with results from simulations. Right-hand panel: probability of contiguous recall as a function of the order of recall. Numerical results displayed correspond to $d=7$, $\kappa=9$. 
 }
\centering 
\end{figure}
     
The left-hand panel of the figure shows the resulting values of the transition probability as a function of the lag. Experimental values have also been included. It can be seen that, in spite of its simplicity, the toy model reproduces qualitatively the lag recency effect for both sets of initial conditions.
 
In the right-hand panel, the probability of contiguous recall has been plotted as a function of time for the two initial conditions; the variable $t$ in the x-axes represents the order of recall. The same behaviour emerges for both sets of experimental conditions: the contiguous recall probability decreases sharply at small times, experiences a minimum shortly after the beginning of the recall process, then increases steadily, all the way to the end of the process. 
 
This is precisely the behaviour we noticed in experimental data (see Fig. 6).  By reasoning in terms of the hierarchical model, we may now try to explain the possible mechanism behind such behavior. 
   
In this model, when a memory has been retrieved, there are two ways it can locate the next memory: by hopping onto another random-walker that is going to locate it, or by continuing to follow the same random-walker that located the first memory. I will refer to the former mechanism as \q{zapping}, to the latter as \q{free association}. 
 
At the beginning of the recall process, zapping dominates: this happens because many unexplored walkers are still present, and the system is testing the possibilities offered by the many available trajectories. Once most random-walkers have been explored, the system focuses on the few left, and is forced to make a greater use of associative recall. 

By construction, zapping is indifferent to serial position; associative recall, on the contrary, favours contiguous memories because each random-walker moves continuously. Therefore, the probability of contiguous recall will begin to grow. 
    
\begin{figure}
\includegraphics[width= \textwidth]{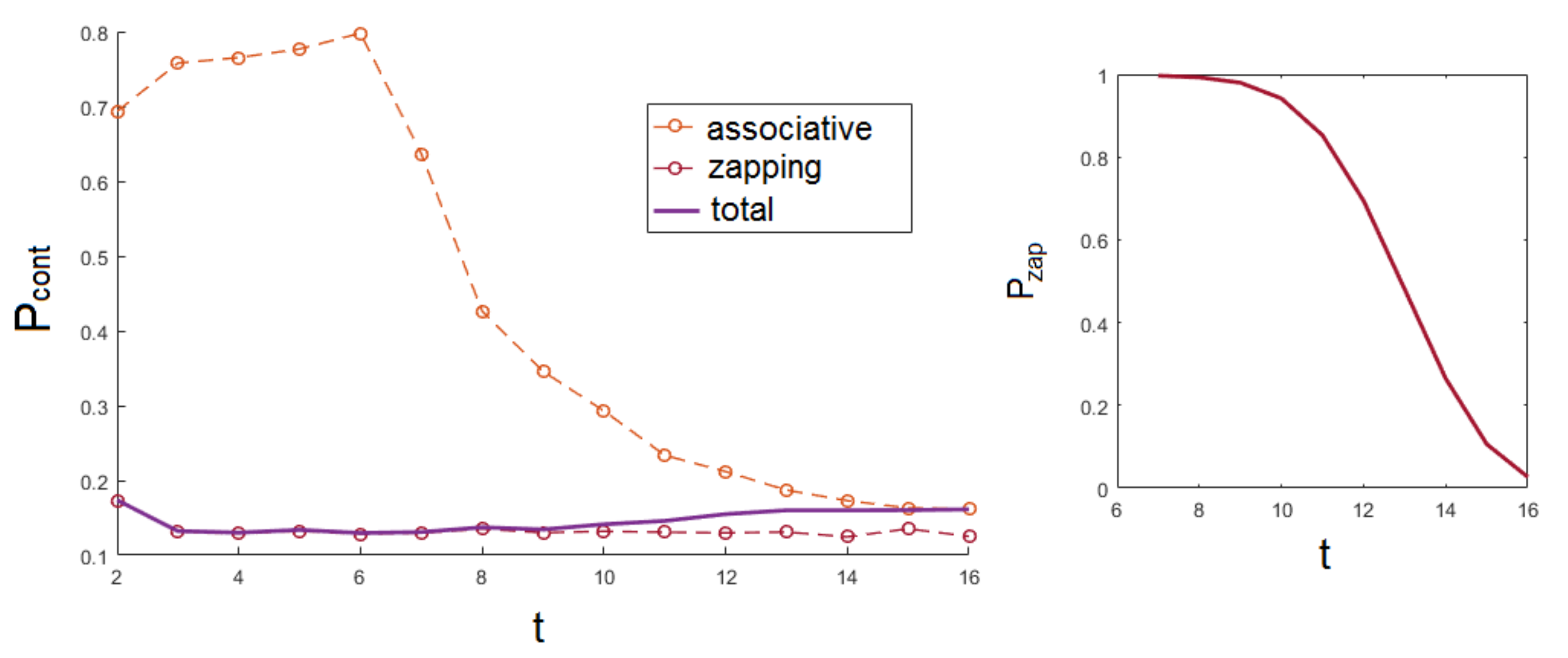}
\caption{Decomposition of the contiguous recall probability in the hierarchical model, with the initial conditions of Healey and Kanaha, 2016. The dashed curves are the rate of contiguity among associative and zapping transitions; the solid curve is the rate of contiguous transitions computed over the full recall process. The right-hand panel shows the time-evolution of the zapping frequency $P_{zap}$. Results correspond to $d=7$, $\kappa=9$.}
\centering 
\end{figure}
    
This interpretation is confirmed by Fig. 9, where the associative component and the zapping component of the contiguous recall probability are shown separately for one of the simulations (with the Heleay and Kahana initial conditions). The observable quantity predicted by the figure is the average probability of contiguous recall, shown as a solid curve, which lies between the two dashed curves corresponding to the zapping and associative components. The smaller panel to the right shows the zapping frequency as a function of time.
   
At small times, zapping predominates. The probability of contiguous transitions coincides therefore with the probability of contiguity among zapping transitions. Since the initial condition is concentrated near few definite locations (the initial and final memories), zapping transitions possess initially a large degree of contiguity. This ceases to be true as the packet of random-walkers spreads away from its initial position. Zapping transitions cease to have a considerable chance of being contiguous, and since they are still the dominant type of transitions, the overall probability of contiguous recall goes down.  
 
Gradually, the zapping frequency decreases and the frequency of associative transitions increases. As a consequence, the average probability of contiguous recall coincides less and less with the probability of contiguous transitions in the zapping process, and migrates toward the associative component. 
  
The lag-recency effect enters now into play. Through its action, as seen in Fig. 9, the curve of contiguous recall for associative transitions lies always above the curve of contiguous recall for zapping transitions. As associative recall comes to dominate the dynamics, the probability of contiguous transitions therefore increases. By the end of the process it virtually coincides with the probability of contiguous transitions in associative recall. 
  
The hierarchical structure of the search for memories leaves its telltale mark in the presence of two different time scales. The initial drop in contiguous recall occurs over a time scale that depends on the dimensionality of psychological space and on the distribution of initial conditions. This is a microscopic time scale, independent on the size of the random-walker population. 
 
The subsequent increase in contiguous recall, on the other hand, happens on a macroscopic time scale, that could not be obtained from studying individual random-walker dynamics. This is the time scale over which the reservoir of unexplored random-walkers is gradually depleted, and its value is controlled by the parameter $\kappa$. We learn from experiments that $\kappa$ is large enough to allow for the depletion to stretch throughout the recall process, but low enough to let the reservoir be substantially depleted. 
  
The rate of contiguous recall has proven to be a highly informative variable. Thinking along similar lines, we may find a qualitative explanation for what we observed in the distribution of the IRIs. 
   
The argument is simple: in the hierarchical scenario, there are two different ways to perform fast transitions between words during the recall process, by exploiting lag-recency or by exploiting zapping. The same mechanism that creates the lag-recency effect can make associative recall faster, but only if the words involved lie nearby within the list. Hence, fast transitions due to association may only occur for small serial-position lags. 
 
Zapping, on the other hand, can provide fast transition with any lags. Going from a random-walker that has found a memory to another random-walker that has just found another memory is equally likely regardless of the locations of the memories, and may produce fast transitions between memories located at any lag from each other. Only for very long lags, however, zapping encounters no rival mechanism, because associative recall performs those transitions too rarely to provide any competition. 

\begin{figure}
\includegraphics[width= \textwidth]{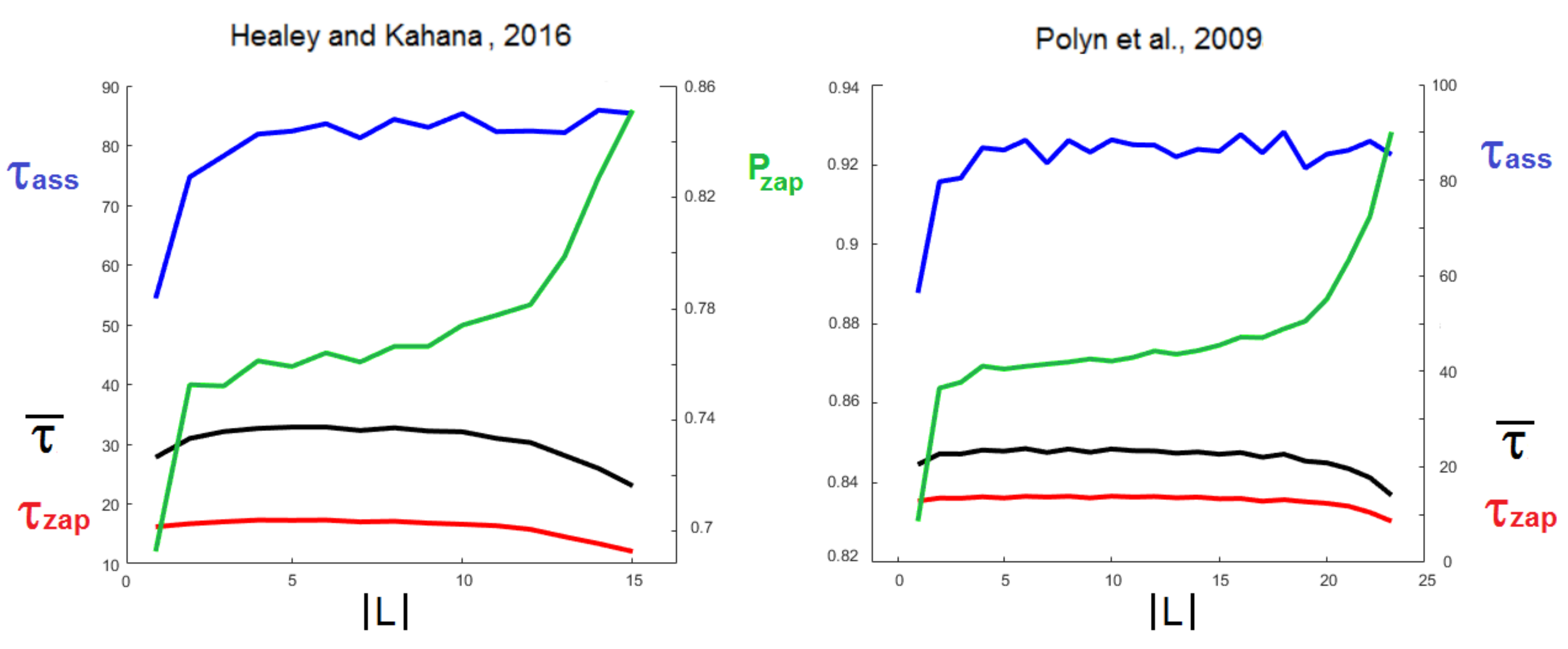}
\caption{Mean inter-response interval $\tau$ for transitions with lag $L$: restricted to associative transitions ($\tau_{ass}$, blue), restricted to zapping transitions ($\tau_{zap}$, red), and unrestricted ($\overline{\tau}$, black). The zapping probability $P_{zap}$ is shown in green. The simulations were performed in $7$ dimensions with $\kappa =9$. Initial conditions were taken from the experiments mentioned in the headings.}
\centering 
\end{figure}

Therefore, two different sets of fast transitions should exist, one from association, at short lags, and one from zapping, at long lags. 
   
To verify this conjecture, I employ simulations of the model with the same parameters as above ($7$ dimensions, $\kappa=9$) so as to extract: (1) the mean IRI; (2) the mean IRI among associative transitions; (3) the mean IRI for zapping transitions; (4) the zapping probability. All this is plotted in Fig. 10 as a function of the lag, for both sets of initial conditions. The resulting scenario appears identical for the two initial conditions, and sheds some light on the possible meaning of Fig. 2. 
 
The time analog of the lag-recency effect speeds up associative transitions only for very short lags; in this range, hence, the mean associative IRI (shown in blue) grows quickly as a function of the lag. 

For such short lags, the zapping probability (shown in green) is considerably smaller than unity, and associative transitions play a crucial role. Therefore the average IRI (the only observable quantity, shown in black) reflects strongly the increase of the associative IRI -- the more so as the zapping IRI (shown in red) is nearly constant. This results in a growth of the mean IRI, as seen in the experiments. 
 
As the lags increases further, the zapping probability experiences a steep growth and associative transitions become rarer. The mean IRI shifts therefore toward the value of the IRI for zapping transitions.  But the zapping IRI lies always beneath the associative IRI (in the whole range of lags); therefore for long lags the mean overall IRI decreases. 
 
As an outcome, the behaviour of the average IRI as a function of the lag corresponds to the black curves of Fig. 10 -- non-monotonous, increasing for short lags and decreasing for long lags -- thus reproducing the experimental results of Fig. 2. 
    
The initial conditions (see Fig.4) produce a further effect clearly discernible in Fig. 10, which adds slightly to the speeding up of transitions with near-maximal lags. Since the random-walker packet is distributed mostly on extremal word memories (near the end and the beginning of the list), zapping transitions with near-maximal lags are \q{prepared} by the initial condition, and happen without much searching at the very beginning of the recall process. That makes them faster, so the zapping IRI also decreases for very long lags. 
   
As we have seen, the concentration of random-walkers on early word-memories is greater for shorter lists, i.e. in the experiments of Healey and Kahana. Accordingly, the decrease of the zapping IRI at long lags is more sustained for those initial conditions. 

A final note: the indentation in the curves of the associative IRI (and, to a lesser extent, in the other curves) does not come from numerical inaccuracy; it is a corollary of the not-so-high dimensionality used for simulations. In the $d$-dimensional space of binary arrays, any two memories can be at most $d$ flips apart. If the lag between them is odd, and the dimension not very large, there is a non-negligeable chance that their distance will be unity, which makes them easier to recall associatively. This indentation, however, gradually disappears in higher dimensions and has no psychological significance. 

\subsection{Mathematical considerations}

The motion of a single random walker in the space of binary arrays of length $d$ is markovian, with transition matrix $M_0 (x, y)= \frac{1}{d} \delta_{{\bf  |x - y|}, 1}$. The associated distribution function $\Psi_x(t) $ (i.e. the probability of finding the walker at position $x$ at time $t$) can be found through the techniques of Kac (1947) and is equal to $\Psi_x(t) = \sum_y {{d} \choose {\|x - y\|}}^{-1}  P_{\|x-y\|}(t) \Psi_y(0)$, where

\begin{equation} 
P_n(t) = \frac{1}{2^{d}} \sum_{i=0}^{d} \  \sum_{k=\max\{ 0 , n-d+i\}}^{\min\{ n, i\}} \frac{(-1)^k d! ( 1 - 2 i/ d)^t  
}{k! (i-k)! (n-k)! (d-i-n+k)!}
\end{equation}
 
In the full stochastic process involving $\kappa S$ random walkers, the position of the random-walker currently entitled to retrieval evolves according to the transition matrix

\begin{equation}
\label{reformulation}
M_t(x,y) = ( 1 - \alpha_t) M_0(x,y) + \alpha_t \Psi _x(t)
\end{equation}

where $\{\alpha_t\}_{t\geq 0}$ is random binary sequence constrained by $\sum_t  \alpha_t = \kappa N  -1$.  For each given choice of the sequence $\alpha_t$, the resulting process is a nonhomogeneous Markov chain, and the stochastic object it describes is a single collective random-walker moving in the random field created by the upper level of the hierarchical search. 

The model discussed in this paper should not be taken as a realistic portrait of the actual search mechanism. It is mainly designed to grasp one nontrivial aspect of the psychology involved -- the coexistence of zapping and free association, with the gradual discarding of random-walkers that causes a dwindling of the zapping phase. 
  
To infer in what ways the real search differs from the basic model I have proposed, a principle of functionality may be invoked. If we accept the above demonstration that multiple thought-particles are being deployed, the actual search mechanism must be designed to utilise them efficiently. 

For instance, random-walkers may only be discarded if they have proven inefficient throughout a certain number of steps. Or, when selecting which random-walkers to visit next, the system may use some approximate knowledge on the current findings of the random-walkers: e.g., only walkers that have just located a memory may compete for retrieval rights. These more realistic models turn out to require far fewer random-walkers and far less time to perform retrieval (compare Alon et al., 2008).

For these models, the reformulation of the theory as the motion of a single collective random-walker, as per eq. (\ref{reformulation}), will no longer be exact. If the macroscopic random-walk uses feedback on the random-walker trajectories, the macroscopic and microscopic motions do not decouple. However, the reformulation will still be valid at a mean-field level, with the hopping probability $\alpha_t$ now solved for self-consistently. 
 
This approach will be further developed elsewhere. Let us only notice that, for the class of models we are considering, the mean-field theory holds a valuable psychological significance. It describes the viewpoint of an observer who interacts with the gas of random-walkers from the outside, unawares of its multiplicity.  

\section{Conclusions} 
     
Two instances of non-monotonicity in free recall have been identified through a fresh analysis of data from two independent experiments. 

The distribution of the inter-response intervals (IRIs) exhibits a peak at low durations for very short lags, and a similar peak for very long lags, while for intermediate lags the peaks tend to be suppressed. As a consequence, the mean and median values of the IRI experience a minimum at intermediate lag values. 
The rate of contiguous recall, on the other hand, exhibits a minimum near the beginning of the recall process. This minimum appears to be robust,  and its position an increasing function of the list's size.

While these findings are counterintuitive, they may be understood by assuming that multiple retrieval processes are being carried out simultaneously, and by allowing the retrieval process to switch stochastically among them.  This leads to a dynamics where memories can be reached either through the switching mechanism (\q{zapping}) or continuously, through free association.
 
To test the hypothesis, I have considered a simple two-parameter model, where multiple random-walkers are allowed to explore psychological space simultaneously. The right to effect retrieval is passed around among random-walkers and can be retained for any length of time, but no random-walker is granted retrieval rights twice. 

This affects drastically the curve of contiguous recall, and produces a behaviour compatible with experiments. In the early stages of the process, transitions between different random-walkers dominate. Later, associative retrieval through single-walker trajectories is prevalent and contiguous recall intensifies accordingly. 
 
The distribution of the IRIs is also given an explanation: the first peak comes from the enhancement of associative retrieval at low lags; the second peak comes from the shorter time interval required by transitions between memories retrieved by two different random-walkers.

I would like to thank Michael J. Kahana, of the University of Pennsylvania, for generously sharing the data obtained in his laboratory. 

\section{Bibliography}
Abbott, J.~T. (2015). Random walks on semantic networks can resemble optimal foraging. \emph{Psychol. Rev.}, \emph{122}(3): 558--69.

Alon N., Avin C., Kouchy M., Kozma G., Lotker Z. \& Tuttle M. (2008). Many random walks are faster than one. In: \emph{Proc. of SPAA 2008}, pp.  119--128. 

Binet A. \& Henri. V. (1894). La memoire des mots. \emph{L'ann\'ee psychologique}, Bd. I, 1:1--23. 
 
Bjork R.~A.  \&  Whitten W.~B. (1974).  Recency-sensitive retrieval processes in long-term free recall. \emph{Cognitive Psychology}, \emph{6}(2): 173--189. 

Bonabeau, E., Theraulaz, G., \&  Dorigo M. (1999). \emph{Swarm intelligence: From Natural to Artificial Systems}. Oxford (UK): Oxford University press. 

Cowan, N. (2010). Multiple Concurrent Thought: the Meaning and Developmental Neuropsychology of Working Memory. \emph{Dev. Neuropsychl.}, \emph{35}(5): 447--474. 

Ebbinghaus, H. (1913). \emph{Memory: A contribution to experimental psychology}. New York, NY: Teachers College, Columbia University.

Eberhart, R.~C., Shi Y., 
\&  
Kennedy J. (2001). \emph{Swarm Intelligence}. New York, NY: Morgann Kaufmann.

Healey \& Kahana(2016). A four-component model of age-related memory change. \emph{Psychological Review}, \emph{123}(1):23--69. 
 
Howard M.~W. \& Kahana M.~J. (2002). A distributed representation of temporal context. \emph{Journal of Mathematical Psychology}, \emph{46}:269--299.
  
Kac M. (1947). Random Walk and the Theory of Brownian Motion. \emph{The American Mathematical Monthly}, \emph{54} (7), 1: 369--391. 

Kahana, M.~J. (1996). Associative retrieval processes in free recall. \emph{Memory and Cognition}, \emph{24}: 103--109.

Kahana M.~J. (2012). \emph{Foundations of Human Memory}. Oxford, UK: University Press.
 
Kenett, Y.~N. (2014). Examining Search Processes in Low and High Creative Individuals with Random Walks. \emph{Front. Hum. Neurosc.}, \emph{8}:407--413.

Lohnas L.~J. \&
Kahana M.~J. (2013). Parametric effects of word frequency effect in memory for mixed frequency lists. \emph{Journal of Experimental Psychology: Learning, Memory, and Cognition}, \emph{39}:1943--1946.

Murdock B.~B. (1960). The immediate retention of unrelated words. \emph{Journal of Experimental Psychology}, \emph{60}:222--234.
  
Murdock B.~B. (1962). The serial position effect of free recall. \emph{Journal of Experimental Psychology}, \emph{64}(5):482--488. 
 
Murray D.~J., Pye C., \& Hockley W.~E. (1976). Standing's power function in long-term memory. \emph{Psychological Research}, \emph{38}(4):319--331.

Polyn S.~M., Norman K.~A., \& Kahana M.~J. (2009). A context maintenance and retrieval model of organizational processes in free recall. \emph{Psychological Review}, \emph{116}:129--156.

Romani S., Pinkoviezky I., Rubin A., Tsodyks M. (2013). Scaling laws of associative memory retrieval. \emph{Neural Computation}, \emph{25}:2523--2544.
 
Sigman and Dehaene (2008). Brain Mechanisms of Serial and Parallel Processing. \emph{Journal of Neuroscience}, \emph{23}(30): 7585--7598. 

Standing L. (1973). Learning 10.000 pictures. \emph{Quarterly Journal of Experimental Psychology}, \emph{25}:207-222.

\end{document}